\documentclass[preprints,article,accept,pdftex,moreauthors]{Definitions/mdpi} 
\firstpage{1} 
\pubvolume{1}
\issuenum{1}
\articlenumber{0}
\pubyear{2022}
\copyrightyear{2022}
\hreflink{https://doi.org/universe8040244} 
\pdfoutput=1

\Title{Nonsingular Black Holes in $4D$ Einstein--Gauss--Bonnet Gravity}

\TitleCitation{Nonsingular Black Holes in $4D$ Einstein--Gauss--Bonnet Gravity}

\Author{Arun Kumar $^{1,2,}$*\orcidA{}, Dharmanand Baboolal $^{3}$ and Sushant G. Ghosh $^{1,3}$\orcidC{}}

\AuthorNames{Arun Kumar, Dharmanand Baboolal and Sushant G. Ghosh}

\AuthorCitation{Kumar, A.; Baboolal, D.; Ghosh, S.G.}

\address{%
$^{1}$ \quad Centre for Theoretical Physics, Jamia Millia Islamia, New Delhi 110025, India; sghosh2@jmi.ac.in (S.G.G.)\\
$^{2}$ \quad Department of Mathematical Science, University of Zululand, Private Bag X1001, \mbox{Kwa-Dlangezwa 3886, South Africa}\\
$^{3}$ \quad Astrophysics and Cosmology Research Unit, School of Mathematics, Statistics and Computer Science, University of KwaZulu-Natal, Private Bag X54001, Durban 4000, South Africa; baboolald@ukzn.ac.za (D.B.)}

\corres{Correspondence: arunbidhan@gmail.com}

\abstract{Recently, several methods have been proposed to regularize a $D \to 4$ limit of Einstein--Gauss--Bonnet (EGB), leading to nontrivial gravitational dynamics in $4D$. We present an exact nonsingular black hole solution in the $4D$ EGB gravity coupled to non-linear electrodynamics and analyze their thermodynamic properties to calculate precise expressions for the black hole mass, temperature, and entropy. Because of the magnetic charge,  the thermodynamic quantities are corrected, and the Hawking--Page phase transition is achievable with diverges of the heat capacity at a larger critical radius $r=r_{+}^{C}$ in comparison to the $5D$ counterpart where the temperature is maximum. Thus, we have a black hole with Cauchy and event horizons, and its evaporation leads to a thermodynamically stable extremal black hole remnant with vanishing temperature, and its size is larger than the $5D$ counterpart. The entropy does not satisfy the usual exact horizon Bekenstein--Hawking area law of general relativity with a logarithmic area correction term.}


\begin{document}
\section{Introduction}
Regular (or nonsingular) black holes are widely considered to resolve the singularity problems, dating back to Bardeen, who gave the first regular black hole model by Bardeen~\cite{Bardeen:1968}, according to whom there are horizons, but there is no singularity.   There has been an enormous advance in the analysis and application of regular black holes, and several exciting papers appeared uncovering properties~\cite{Ansoldi:2008jw,Hayward:2005gi,Bronnikov:2000vy,Lemos:2011dq,Schee:2015nua}. Hayward~\cite{Hayward:2005gi} proposed that Bardeen-like, regular space-times describe the formation of a black hole from an initial vacuum region with a  finite density and pressures, vanishing rapidly at a large distance and behaving as a cosmological constant at a small distance. 
The spherically symmetric Hayward's metric~\cite{Hayward:2005gi} is given by
\begin{equation}\label{mtrc}
	ds^2 = -f(r)dt^2+\frac{1}{f(r)}dr^2 +r^2 \left( d\theta^2 +\sin^2 \theta d\phi^2 \right),
\end{equation}
with 
\begin{equation*}
f(r) = 1-\frac{2mr^2}{r^3+2l^2m}. 
\end{equation*}
Here $m$ is mass and $l$ is constant. 
The solution (\ref{mtrc}), for $l=0$, encompasses well-known Schwarzschild black holes and the flat Minkowski spacetime for $m=0$. The regularity of the solution (\ref{mtrc}) is confirmed by the curvature invariants, which are well behaved everywhere, including at $r=0$. 
The analysis $f(r)=0$, implies a critical mass $m^* = 3\sqrt{3}l/4$ and critical radius $r^* = \sqrt{3}l$, such that a regular extremal black hole with degenerate horizons $r=r^*$ when $m=m^*$. When $m<m^*$, a regular non extremal black hole horizon with both Cauchy and event horizons, corresponding to two roots of $f(r)=0$, and no black hole when $m>m^*$~\cite{Hayward:2005gi}.  Further, it is shown that the event horizon is located near $2m$, while the inner one is close to $l$~\cite{Frolov:2017rjz,Frolov:2016pav}. 
Thus the metric (\ref{mtrc}) is a regular black hole, which asymptotically behaves as 
\begin{equation*}
f(r) \sim 1-\frac{2m}{r} \;\;\; \text{as} \;\;\; r \rightarrow \infty, 
\end{equation*} 
where near center 
\begin{equation*}
f(r) \sim 1-\frac{r^2}{l^2} \;\;\; \text{as} \;\;\; r \rightarrow 0.
\end{equation*}  
which implies that at a large 
distance ($r$) it reproduces the Schwarzschild metric, while at the origin, it is not only regular but also has de Sitter form.  
The various properties mentioned above make the metric (\ref{mtrc}) simple for the analysis of its scaling behavior~\cite{Frolov:2016pav}. In addition, the Hayward black hole can be shown as an exact model of general relativity coupled to nonlinear electrodynamics and hence attracted significant attention in various studies, such as  Quasinormal  modes  of the black holes Lin by {et al.}~\cite{Lin:2013ofa}, the geodesic equation of a particle by Chiba and Kimura~\cite{Chiba:2017nml}, wormholes from the regular black hole~\cite{Kuhfittig:2013jna,Halilsoy:2013iza} with  their stability ~\cite{Sharif:2016wvq}, black hole thermodynamics~\cite{Kumar:2018vsm,Maluf:2018lyu} and related properties~\cite{Mehdipour:2016vxh, Perez-Roman:2018hfy}, and strong deflection lensing~\cite{Zhao:2017cwk}.  The rotating regular Hayward metric~\cite{Gwak:2017zwm, Amir:2016nti, Amir:2015pja,Abdujabbarov:2016hnw} and Hayward black hole solution in EGB has been discussed in~\cite{Kumar:2020bqf,Ghosh:2020tgy}.   

Lovelock's theory of gravitation also resembles string inspired models of gravity as its action contains, among others, the quadratic Gauss--Bonnet term. It is the most natural generalization of Einstein's theory to higher dimensions. It is the only Lagrangian-based theory of gravity that gives covariant, conserved, second-order field equations and yields non-trivial dynamics in  $D\geq 5$. This quadratic term naturally arises in the low energy effective action of heterotic string theory~\cite{Gross:1986mw,Bento:1995qc} and it also appears in six-dimensional Calabi-Yau compactifications of M-theory~\cite{Guica:2005ig}. The theory is free of ghosts about other exact backgrounds~\cite{Boulware:1985wk}. EGB gravity~\cite{Lanczos:1938sf} is a particular case of Lovelock theory of gravitation~\cite{Lovelock:1971yv}, whose Lagrangian contains just the first three terms and is of particular interest. It appears naturally in heterotic string theory's low energy effective action~\cite{Gross:1986mw,Bento:1995qc}. Boulware and Deser~\cite{Boulware:1985wk} first discovered the static spherically symmetric black hole, generalizing the Schwarzschild solution in EGB theory to show that the only stable solution, the central singularity, is still unpreventable, which is true for the charged analogous black holes as well~\cite{Wiltshire:1985us}. Later several attractive solutions were obtained for the EGB theory from various sources~\cite{Wheeler:1985nh,Mignemi:1992nt,Kleihaus:2011tg,Ghosh:2011ad,Hansraj:2019hxh,Alexeev:1996vs,Kanti:1996gs}. In the Einstein-scalar--Gauss--Bonnet theory with a coupling function, the no-hair theorems are easily evaded, and also lead to a large number of regular black-hole solutions~\cite{Antoniou:2017acq,Bakopoulos:2018nui,Antoniou:2017hxj} and  other regular black holes are solution Einstein--Gauss--Bonnet theory coupled to nonlinear electrodynamics~\cite{Ghosh1:2018bxg,Hyun:2019gfz,Singh:2019wpu}.  

The case 4-dimensional ($4D$) is special because the Euler--Gauss--Bonnet term becomes a topological invariant that does not contribute to the equations of motion or to the gravitational dynamics. However, a $4D$ EGB gravity is a formulation in which the Gauss--Bonnet coupling has been rescaled as $\alpha/(D-4)$. The $4D$ EGB theory is defined as the limit $D \rightarrow 4$, which preserves the number of degrees of freedom and is thereby free from the Ostrogradsky instability~\cite{Glavan:2019inb}.  Further,  this extension of  Einstein's gravity bypasses all conditions of Lovelock's theorem~\cite{Lovelock:1972vz} and is also free from the singularity problem.  Such $4D$ black hole solutions have been obtained  in the semi-classical Einstein's equations~\cite{Cai:2009ua}, gravity theories with quantum corrections~\cite{Cognola:2013fva}, and  also recently in Lovelock gravity~\cite{Casalino:2020kbt}. A cascade of subsequent interesting work analyzed black hole solutions in the  $4D$ EGB theory, this~includes static, spherically symmetric  Reissner$-$Nordstrom-like solution~\cite{Fernandes:2020rpa}, black~hole surrounded by clouds of string~\cite{Singh:2020nwo}, Vaidya-like solution\cite{Ghosh:2020vpc} and generating  black hole solutions~\cite{Ghosh:2020syx}, noncommutative inspired black holes~\cite{Ghosh:2020cob}, regular black holes~\cite{Kumar:2020uyz}, rotating black holes and their shadows~\cite{Wei:2020ght,Kumar:2020owy}. Furthermore, the quasinormal modes, stability~\cite{Konoplya:2020bxa,Guo:2020zmf,Konoplya:2020juj,Churilova:2020aca}, relativistic~stars solution~\cite{Doneva:2020ped}, the motion of a classical spinning test particle~\cite{Zhang:2020qew}, gravitational lensing~\cite{Islam:2020xmy,Heydari-Fard:2020sib,Jin:2020emq,Kumar:2020sag}, and derivation of regularized field equations~\cite{Fernandes:2020nbq}  were also  explored. 

The original idea of the $4D$ regularization procedure of EGB gravity was given by Tomozawa~\cite{Tomozawa:2011gp}  and  later  Cognola {et al.}~\cite{Cognola:2013fva} simplified the approach by reformulating it within a classical Lagrangian approach. However, several questions were raised ~\cite{Ai:2020peo,Hennigar:2020lsl,Shu:2020cjw,Gurses:2020ofy,Mahapatra:2020rds} on  the $4D$ regularization procedure techniques of~\cite{Glavan:2019inb}  and there were several  alternate proposals for the $4D$  regularization of EGB gravity~\cite{Lu:2020iav,Kobayashi:2020wqy,Hennigar:2020lsl,Casalino:2020kbt,Ma:2020ufk,Arrechea:2020evj,Aoki:2020lig}.  Amongst these,  L\"{u} and Pang ~\cite{Lu:2020iav} regularized the $4D$  EGB gravity, via the Kaluza--Klein-like technique which is a  well-defined and finite action of special scalar-tensor theory that belongs to the family of Horndeski  gravity,  in agreement with the results of Ref.~\cite{Kobayashi:2020wqy}.  Thereafter, \mbox{Hennigar {et al.}~\cite{Hennigar:2020lsl}} proposed another well-defined $D \to 4$ limit of EGB gravity  and this regularization  is applicable not only in $4D$ but also in lower dimensions.  Another alternate method of regularization for $4D$ EGB gravity ~\cite{Fernandes:2020nbq} is proposed  by adding counter terms, sufficient to cancel  divergence of the action, yielding a set of field equations that can be written in closed form for $4D$ spacetime.  However, It turns out that  the maximally symmetric or spherically symmetric $4D$  black hole solution obtained in ~\cite{Glavan:2019inb} remains valid for these alternate  regularized theories~\cite{Lu:2020iav,Hennigar:2020lsl,Casalino:2020kbt,Fernandes:2020nbq,Ma:2020ufk}. 

This paper searches for a static, spherically symmetric nonsingular solution of the $4D$ EGB gravity, namely, whose charge is described by nonlinear electrodynamics (NED), $4D$~nonsingular EGB black holes. Thus, we investigate a  black hole solution in the $4D$ EGB gravity coupled to a NED theory. We study not only the structure of $4D$ nonsingular EGB black holes but also its thermodynamic properties, including the stability of the system. In~particular, we explicitly bring out how the effect of NED can alter the black hole solution and its thermodynamic properties. 

\section{Basic Equations and Black Hole Solution}
\label{BAdS_section}
We consider fully interacting theory of gravity  minimally coupled to nonlinear electrodynamics (NED), whose action in $D$-dimensional spacetime is given by~\cite{Kumar:2018vsm,Ghosh1:2018bxg}
\begin{equation}\label{action}
\mathcal{I}=\int d^{D}x\sqrt{-g}\mathcal{L}\equiv\mathcal{I}_{gravity}+\mathcal{I}_{NED}.
\end{equation} 
The gravity action with the metric $g_{ab}$ is given by
\cite{Cai:2001dz,Dehghani:2006ke,Ghosh:2020tgy}
\begin{equation}
\mathcal{I}_{gravity}=\frac{1}{16\pi}\int d^{D}x\sqrt{-g}\left[R+\alpha\mathcal{L}_{GB}\right],
\end{equation}
where $R$ is Ricci scalar. The action (\ref{action}) is the modification of the Einstein--Hilbert action with a quadratic curvature correction Gauss--Bonnet (GB) term, and $\alpha$ is the GB coupling constant of dimensions $[length]^2$ and
\begin{equation}
\mathcal{L}_{GB}=R^2- 4 R_{ab}R^{ab}+ R_{abcd} R^{abcd}.\nonumber
\end{equation}
The NED action~\cite{Dehghani:2006ke,Ghosh:2020tgy}
\begin{equation}
\mathcal{I}_{NED}=\frac{1}{16\pi}\int d^{D}x \sqrt{-g} \mathcal{L}(F),
\end{equation}
with $F=F_{ab}F^{ab}/4$ where $F_{ab}$ is a filed strength tensor, such that $F_{ab}=\partial_{a}A_{b}-\partial_{b}A_{a}$. $A_{a}$ is the gauge potential with corresponding tensor field $\mathcal{L}(F)$. Varying the action (\ref{action}), we obtain the following equations of motion~\cite{Dehghani:2006ke,Cai:2001dz}
\begin{equation}
G_{ab}+\alpha H_{ab}=T_{ab}\equiv 2\left[\frac{\partial \mathcal{L}(F)}{\partial F} F_{ac}F^{c}_{b}-\frac{1}{4}g_{ab}\mathcal{L}(F) \right],\label{FieldEq}
\end{equation}
where $T_{ab}$ is the energy momentum tensor for the NED field. Einstein tensor $G_{ab}$ and Lanczos tensor $H_{ab}$~\cite{Lanczos:1938sf,Kastor:2006vw} are
\begin{equation}
\begin{array}{lcl}
G_{ab}&=&R_{ab}-\frac{1}{2}R g_{ab}, \vspace{6pt}\\
H_{ab}&=&2\Bigr( R R_{ab}-2R_{ac} {R}{^c}_{b} -2 R_{acbd}{R}^{cd} - R_{acde}{R}^{cde}{_b}\Bigl)-\frac{1}{2}\mathcal{L}_{GB}g_{ab}.
\end{array}
\end{equation}
 The tensor $F_{ab}$ obeys
\begin{equation}\label{ee1}
\nabla_{a}\left(\frac{\partial \mathcal{L}(F)}{\partial F}F_{ab}\right)=0
\end{equation}
and the Bianchi identities
\begin{equation}\label{dual}
\nabla_{a}\left(^*F_{ab}\right)=0,
\end{equation}
where $*$ denotes the Hodge dual. 
The Maxwell field tensor in $D \geq 5$ reads~\cite{Ghosh:2020tgy}
\begin{equation}
F_{ab}=2\delta^{\theta_{D-3}}_{[a}\delta^{\theta_{D-2}}_{b]}\frac{{g(r)}^{D-3}}{r^{D-4}}\sin\theta_{D-3}\left[\prod_{j=1}^{D-4}\sin^2\theta_{j}\right],
\end{equation}
and then $ F$ becomes
\begin{equation}\label{ff}
 F=\frac{g^{2(D-3)}}{2r^{2(D-2)}}
\end{equation}
The nonsingular solution we are interested in can be derived  from the following $D$ dimensional Lagrangian density of NED  field~\cite{Ghosh:2020tgy}
\begin{equation}\label{lf}
\mathcal{L}(F) =\frac{2(D-1)(2g^2F)^{\frac{D-1}{D-2}}}{sg^2(1+(\sqrt{2gF})^{\frac{D-1}{D-2}})}\quad \text{with} \quad s=\frac{ g^{D-3}}{(D-2)\mu}.
\end{equation} 
Using Equation (\ref{ff}) in Equation (\ref{lf}), we get
\begin{equation}
\mathcal{L}(F)=\frac{2(D-1)(D-2)\mu g^{D-1}}{(r^{D-1}+g^{D-1})^2}.
\end{equation}
The $D$-dimensional static, spherically symmetric metric anstaz reads as~\cite{Kumar:2018vsm,Ghosh:2020tgy}
\begin{equation}
ds^2=-f(r)dt^2+f(r)^{-1}dr^2+r^2d\Omega_{D-2}^2,\label{metric},
\end{equation} 
with $f(r)$ is the metric function to be determined and
\begin{equation}
d\Omega^2_{D-2}=d\theta_1^2+\sum_{i=2}^{D-2}\left[\prod_{j=2}^i \sin^2\theta_{j-1} \right]d\theta^2_i,
\end{equation} 
is the line element of a $(D-2)$-dimensional unit sphere~\cite{Myers:1986un,Xu:1988ju}.  By using metric (\ref{metric}) in Equation \eqref{FieldEq}, by re-scaled coupling constant $\alpha/(D-4)$, we obtained the Einstein--Gauss--Bonnet gravity equations and then the (\emph{r,r}). equation of motion~\cite{Ghosh:2014pga}, which in the limit $D\to4$, takes the following form
\begin{equation}
r^3f^{\prime}(r)+ \alpha\big(f(r)-1\big)\Big(f(r)-1-2rf^{\prime}(r)\Big) + r^2\big(f(r)-1\big)= -\frac{12\mu g^3 r^4}{(r^3+g^3)^{2}},\label{rr1}
\end{equation} 
whose integration leads us to the solution
\begin{equation}
f_{\pm}(r)=1+\frac{r^2}{2\alpha}\left(1\pm\sqrt{1+\frac{8M\alpha}{r^3+g^3}}\right),\label{fr}
\end{equation} 
where $M$ is the constant of integration which is identified as the mass of the black hole and it is related to $\mu$ via $M=2\mu$. The $\pm$ sign in front of the square root term in Equation \eqref{fr}, corresponds to two different branches of solution. The two  branches of the  solution (\ref{fr}), in~the limit $\alpha \to 0$ or GR limit, behave as 
\begin{equation}
\begin{array}{l}
f_{-}(r)=1-\frac{2Mr^2}{r^3+g^3}, \vspace{6pt}\\
f_{+}({r})=1+\frac{2Mr^2}{r^3+g^3}+\frac{r^2}{\alpha}.
\end{array}
\end{equation}
Thus, the $-$ve branch corresponds to the Hayward solution (\ref{mtrc}) ($g^3=2ml^2$) with positive gravitational mass, whereas the $+$ve branch reduces to the Hayward ds/AdS with negative gravitational mass. The magnetic charge $g$ is indeed related to the length $l$, which is a parameter of the length $l$ associated with the region concentrating the central energy density. Modifications in the spacetime metric appear when the curvature scalar becomes comparable with $l^{-2}$~\cite{Frolov:2016pav}. Moreover, the nonzero value of $l$ prevents the curvature scalars from growing infinitely at the central region and makes them bounded from above, just in the spirit of the original idea of the regular black hole~\cite{Kumar:2019pjp}. The regular solution (\ref{fr}), besides critical scale parameter $l$, corresponding to UV complete theory, also contains such parameters as mass $M$ and charge $g$, which specify the solutions. Here, the regularity means that for a fixed value of these black hole parameters, the curvature of the spacetime is finite~\cite{Frolov:2016pav,Kumar:2019pjp}. The authors~\cite{Kumar:2020yem} use the EHT observation of the M87* black hole shadow to investigate the constraints when rotating regular black holes (non-Kerr) can qualify as astrophysical black hole candidates. Indeed, the shadows of Hayward black holes ($g \leq 0.65 M$) are indistinguishable from Kerr black hole shadows within the current observational uncertainties, and thereby they can be viable solid candidates for the astrophysical black holes. The monopole mass has been considered a parameter, and direct monopole production has been performed in most accelerators. The lack of monopole detection has been transformed into a monopole mass lower bound of 400 GeV~\cite{CDF:2005cvf}.  

In an asymptotic large $r\gg g$ limit, the two branches asymptotically take the form
\begin{equation}
\begin{array}{l}
\lim_{r\to \infty}f_+(r)=1+\frac{2M}{r} +\frac{r^2}{\alpha}+\mathcal{O}\Big(\frac{1}{r^4}\Big),\vspace{6pt}\\
\lim_{r\to \infty}f_-(r)=1-\frac{2M}{r} +\mathcal{O}\Big(\frac{1}{r^4}\Big).
\end{array} 
\end{equation} 
Now, the $-$ve branch corresponds to the Schwarzschild black hole solution, on the other hand, the $+$ve branch does correspond to Schwarzschild dS/AdS black hole with negative mass and not physical. Thus, we shall only consider the $-$ve branch of solution (\ref{fr}) which can be identified as a static spherically symmetric Hayward-like regular black hole in the $4D$ EGB gravity, which reduces to the one in Ref.~\cite{Glavan:2019inb} when one switches off the magnetic monopole charge ($g=0$). Henceforth, we can refer to solution (\ref{fr}) as the $4D$ Hayward-EGB black holes or $4D$ nonsingular EGB black holes. 
In the limit $M=0$, $4D$, the nonsingular EGB black holes solution becomes flat as
$f(r)=1$, and near origin it takes the form
\begin{equation}
f(r)=1+\frac{r^2}{l^2_{eff}} \quad \text{with} \quad  \frac{1}{l_{eff}^2}=\frac{1}{2\alpha}\left(1-\sqrt{1+\frac{8M \alpha}{g^3}}\right),
\end{equation} 
and behaves like de Sitter.

The weak energy condition states that $T_{ab}t^at^b\geq 0$ for all time like vectors $t^a$~\cite{se}, i.e., for~any observer, the local energy density must not be negative. 
Hence, the energy conditions require $\rho\geq 0$ and $\rho+P_i\geq 0$, with $P_i=-\rho-\frac{r}{2} \rho'$
\begin{equation}
\begin{array}{r}
\label{ec}
\rho=\frac{3 g^2M }{(r^3+e^3)^{2}}\vspace{6pt}\\
\rho+P_2=\rho+P_3=\rho+P_4=\frac{9g^2r^3 M}{(r^3+g^3)^{3}}.
\end{array}
\end{equation}
Thus, one can notice that the $4D$ nonsingular EGB black hole satisfies the weak energy conditions.

\subsection*{Horizon Structure}
\begin{figure}
\includegraphics[width= 7 cm]{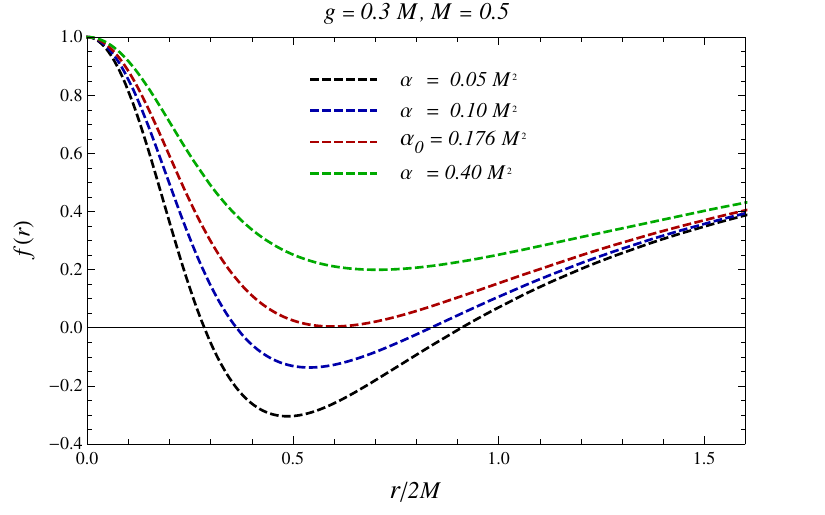}
\includegraphics[width= 7 cm]{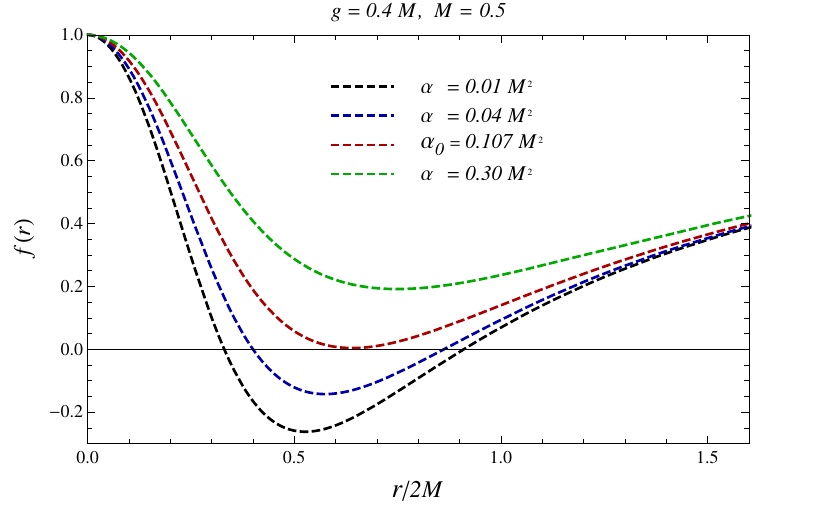}\\
\includegraphics[width= 7 cm]{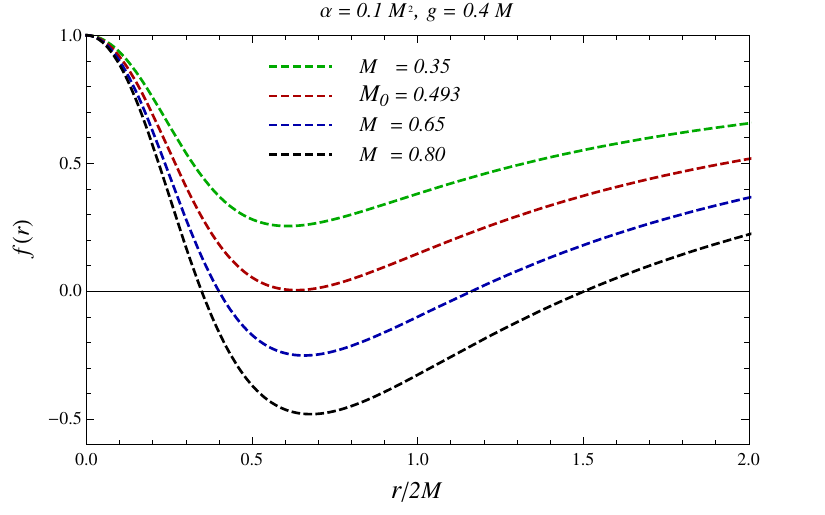}
\includegraphics[width= 7 cm]{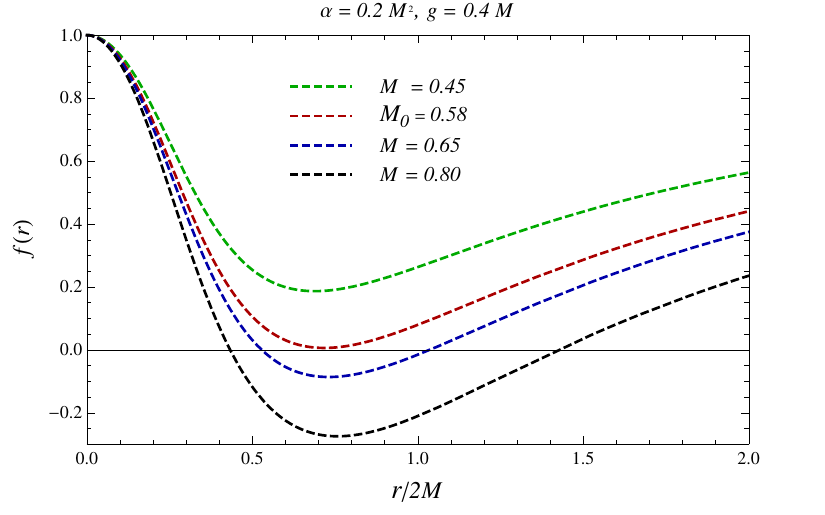}\\
\includegraphics[width= 7 cm]{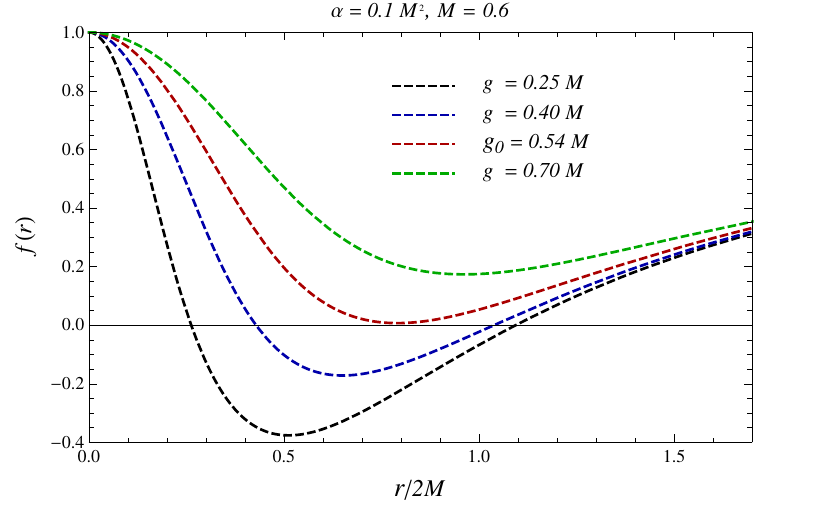}
\includegraphics[width= 7 cm]{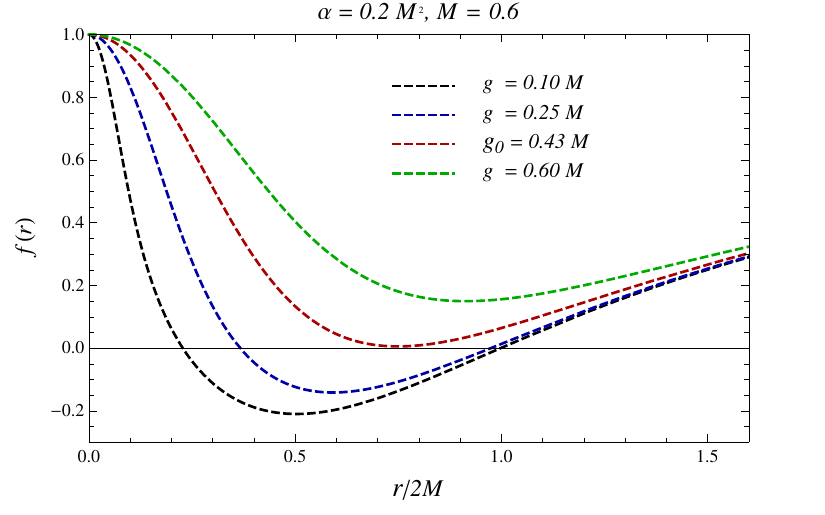}
\caption{ The horizons of the $4D$ nonsingular EGB black holes for various values of $\alpha$, $M$ and $g$.}
\label{fig:f1}
\end{figure}

Next, we discuss the horizon for $4D$ nonsingular EGB black holes. The coordinate singularity of the metric Equation~(\ref{metric}) at $f(r)=0$, implies that the black hole horizons exist. Thus, for the given values of $g$ and $\alpha$, the radii of the horizons are the zeros of 
\begin{equation}
(r_H^3+g^3)(r_H^2+\alpha)-2Mr_H^4=0.\label{horizon}
\end{equation}
By switching off magnetic charge ($g=0$), one gets~\cite{Cho:2002hq,Ghosh:2020tgy} 
\begin{equation}
r^2_H+-2M r_H+\alpha=0.\label{horizon1}
\end{equation}
which admits the following solutions
\begin{equation}
r_{\pm}=M\pm \sqrt{M^2-a},
\end{equation}
where $r_-$ and $r_+$ are, respectively, representing the Cauchy and the event horizons.
The~analysis of Equation \eqref{horizon} leads us to find, for a given set of values $g$ and  $M$, a maximum allowed value of the Gauss--Bonnet coupling constant $\alpha= \alpha_0$ (cf. Figure~\ref{fig:f1}) such that for $\alpha<\alpha_0$, we obtain a black hole solution with double horizons, say $r_{\pm}$, where $r_-$  and $r_+$, respectively, denote the Cauchy and the event horizon. For $\alpha=\alpha_0$, we obtain an extremal black hole solution with degenerate horizon $r_+=r_-\equiv r_E$. One can also find a maximum (or minimum) allowed value $g=g_0$ (or $M=M_0$) for the given values of $M$ and $\alpha$ (or $g$ and $\alpha$). From Figure~\ref{fig:f1}, it is evident that the maximum allowed value of $\alpha$ (or $g$) decreases as we increase the value of $g$ (or $\alpha$), whereas, the minimum allowed value of $M$ increases with $\alpha$. We summarize the Cauchy and the event horizon radii of $4D$ nonsingular EGB black holes for various values of $\alpha$, $g$ and $M$, in Table~\ref{tahor}. It is evident that the Cauchy (event) horizon radius increases (decreases) as we increase $\alpha$ and $g$ (cf. Table~\ref{tahor}), but, this trend reverses for increasing $M$.
\begin{table}[H]
\caption{Cauchy horizon radius ($r_-$), the event horizon radius ($r_+$) and $\delta = r_+-r_-$ of $4D$ nonsingular EGB black hole for different values of parameters.\label{tahor}}
\newcolumntype{C}{>{\centering\arraybackslash}X} 
\begin{tabularx}{\textwidth}{CCCC}
		\toprule
		\textbf{$M = 0.5,~~g = 0.3$}\\
	\midrule
			\textbf{$\alpha$}	& \textbf{$r_-$}	& \textbf{$r_+$}     & \textbf{$\delta$} \\
			\midrule
			0.05 & 0.288  & 0.919 & 0.631\\
			0.10 & 0.358  & 0.843 & 0.485\\
			$\alpha_0$=0.176 & ~0.597  & 0.597 & 0\\
		\midrule
		\textbf{$M = 0.5,~~\alpha = 0.1$}\\
	\midrule
			\textbf{$g$}	& \textbf{$r_-$}	& \textbf{$r_+$}     & \textbf{$\delta$} \\
			\midrule
			0.25 & 0.268  & 1.107 & 0.839\\
			0.40 & 0.432  & 1.044 & 0.612\\
			$g_0$=0.54 & 0.797  & 0.797 & 0\\
			\midrule
			\textbf{$g = 0.4,~~\alpha = 0.1$}\\
		\midrule
			\textbf{$M$}	& \textbf{$r_-$}	& \textbf{$r_+$}     & \textbf{$\delta$} \\
			\midrule
			0.80 & 0.352  & 1.513 & 1.161\\
			0.65 & 0.406  & 1.172 & 0.766\\
			$M_0$=0.493 & 0.6515  & 0.6515 & 0\\	
			\bottomrule
		\end{tabularx}
\end{table}

\section{ Thermodynamics}
\label{sec3}
The conclusion by Wheeler~\cite{Ruffini:1971bza} that any system consists of a black hole violates the non-decreasing entropy law necessitated to assign temperature and entropy to a black hole. This association of thermodynamic quantities with black holes led the study of black holes as a thermodynamic system. The black hole mass, in terms of the horizon radius $r_+$ by solving equation $f(r_+)=0$, reads
\begin{equation} \label{B_mass}
M_+= \frac{r_+}{2} \left[\left(1+\frac{\alpha}{r_+^2}\right)\left(1+\frac{g^3}{r_+^3}\right)\right],
\end{equation}
by taking $g=0$, one obtains mass for $4D$ EGB black holes~\cite{Glavan:2019inb,Fernandes:2020rpa,Ghosh:2020tgy,Singh:2020nwo}
\begin{equation} \label{B_mass1}
M_+= \frac{r_+}{2} \left[1+\frac{\alpha}{r_+^2}\right],
\end{equation}
which further goes over to the mass of Schwarzschild black holes~\cite{Cho:2002hq,Kumar:2018vsm} in GR limits ($\alpha \to 0$), we have $ M_+= {r_+}/{2}.$

The Hawking temperature associated with horizon radius $r_+$, can be obtained through the relation, $T_+=\kappa/2\pi$~\cite{Bardeen:1973gs}, where $$\kappa^2=-\frac{1}{2}\bigtriangledown_{\mu}\chi_{\nu}\bigtriangledown^{\mu}\chi^{\nu}$$ is the surface gravity. Then the  temperature, on using  (\ref{fr}), yields
\begin{equation} \label{B_temp0}
T_+= \frac{1}{4\pi r_+}\left[\frac{r_+^2-\alpha-2\frac{g^3}{r_+^3}\left(r_+^2+2\alpha\right)}{\left(1+\frac{g^3}{r_+^3}\right)\left(r_+^2+2\alpha\right)}\right].
\end{equation}
One gets the expression of temperature for $4D$ EGB black holes~\cite{Glavan:2019inb,Ghosh:2020tgy,Singh:2020nwo,Fernandes:2020rpa}, by keeping $g=0$ 
\begin{equation} \label{B_temp1}
T_+= \frac{1}{4\pi r_+}\left[\frac{r_+^2-\alpha}{r_+^2+2\alpha}\right],
\end{equation}
 further, in GR limits ($\alpha\to 0$), we obtain the temperature of Schwarzschild black holes~\cite{Cho:2002hq,Kumar:2018vsm}
\begin{equation}
T_+= \frac{1}{4\pi r_+}.
\end{equation}

\begin{figure}[H]
\includegraphics[width=7 cm]{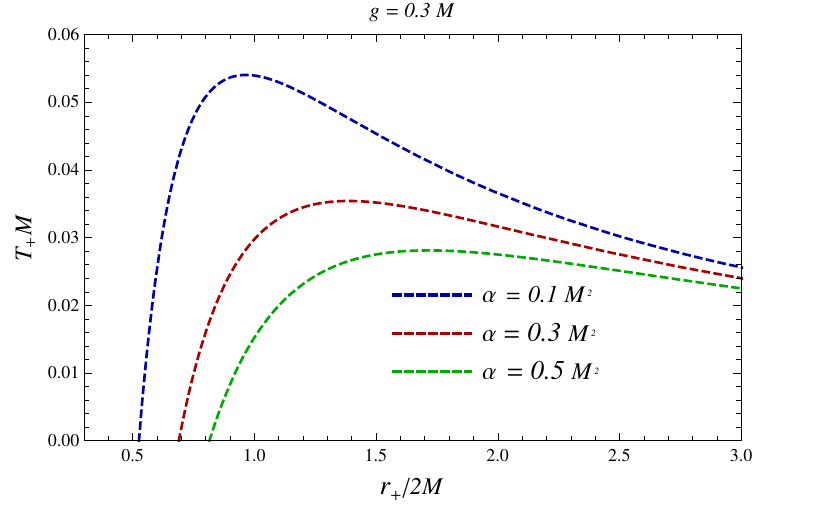}
\includegraphics[width=7 cm]{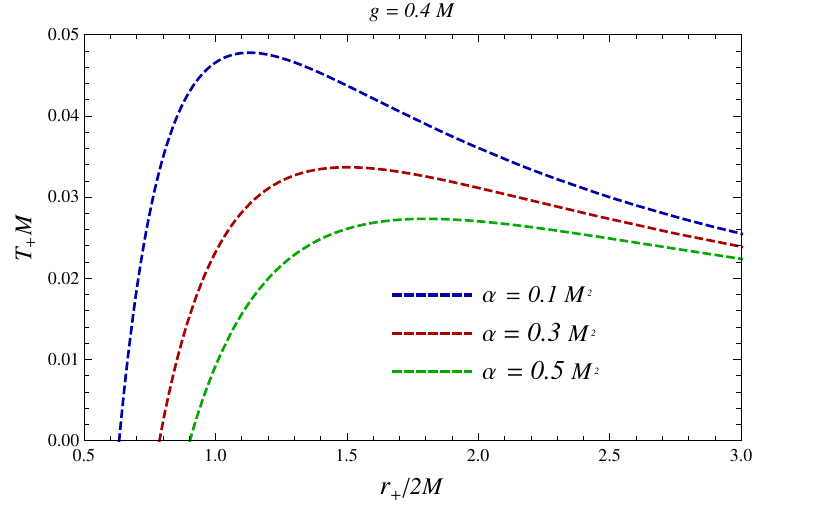}\\
\includegraphics[width=7 cm]{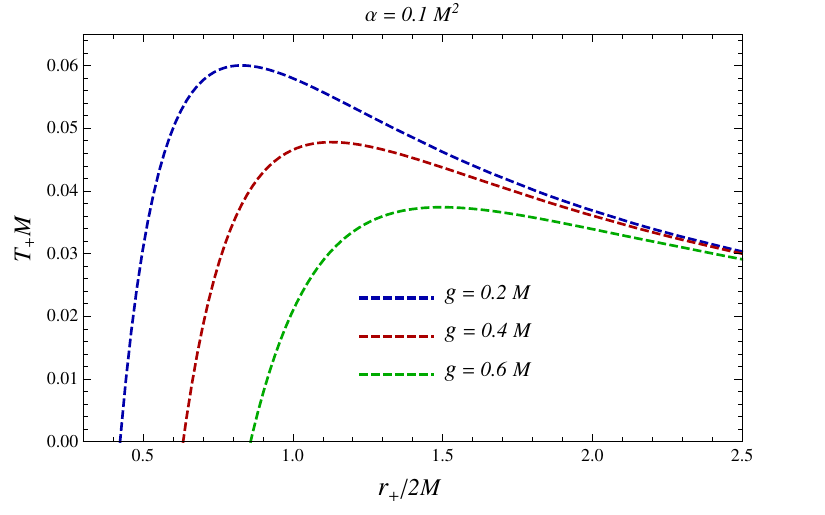}
\includegraphics[width=7 cm]{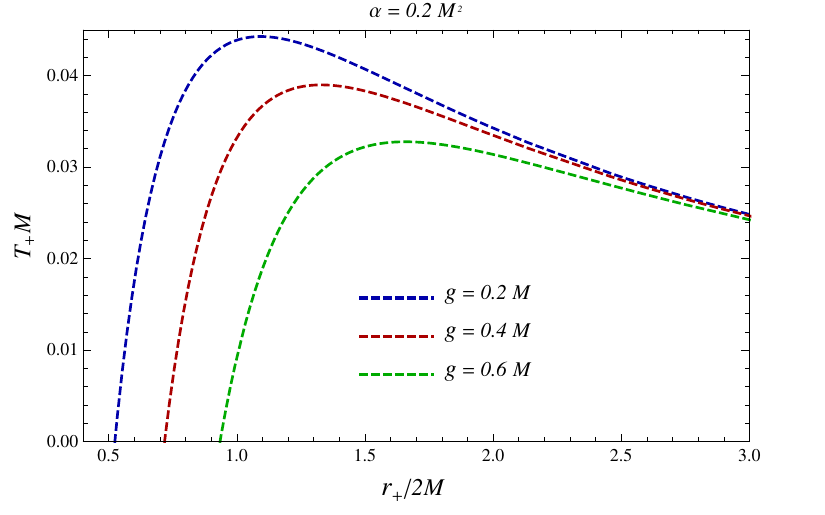}
\caption{$4D$ nonsingular EGB black holes Hawking temperature ${T_+}$ \textit{vs} horizon $r_+$ of  for various values of $\alpha$ and $g$.}
\label{fig:Btemp1}
\end{figure}

We analyze the behavior of the Hawking temperature of $4D$ nonsingular EGB black hole through temperature $T_+$ \text{vs} horizon radius $r_+$ plots shown in Figure~\ref{fig:Btemp1}. When the $4D$ nonsingular EGB black hole undergoes Hawking evaporation, its temperature increases as the event horizon shrinks to reach a maximum value before decreasing rapidly to vanish at some particular value of horizon radius. One can notice that the Hawking temperature, at a particular value of horizon radius, $r_+^c$ (critical radius), possesses a local maximum, which means the first derivative of temperature vanishes leading to the divergence of specific heat. The numerical results of critical radius $r_+^c$ with corresponding maximum temperature $T_+^{Max}$ have been summarized in Table~\ref{tr3}, from which one can notice that an increment in magnetic monopole charge $g$ or the Gauss--Bonnet coupling constant $\alpha$ results in the increased and decreased values of $r_+^c$ and $T_+^{Max}$, respectively. Further, when we compared the temperature of $5D$ nonsingular EGB black holes~\cite{Kumar:2020bqf}, the local maximum in the $4D$ case occurs at a larger horizon radius.

\vspace{-6pt}

\begin{table}[H]
\caption{The critical radius ($r_+^c$) with corresponding maximum temperature ($T_+^{Max}$) of $4D$ nonsingular EGB black hole for different values of $\alpha$ and $g$.\label{tr3}}
\newcolumntype{C}{>{\centering\arraybackslash}X}
\begin{tabularx}{\textwidth}{CCC}
		\toprule
			\textbf{$g = 0.3$}\\
		\midrule
			\textbf{$\alpha$}	& \textbf{$r_+^C$}	& \textbf{$T_+^{Max}$} \\
			\midrule
			0.1 & 0.9656  & 0.054\\
			0.3 & 1.3862  & 0.0353\\
			0.5 & 1.7165  & 0.0283\\
		\midrule
		\textbf{$\alpha = 0.1$}\\
		\midrule
		\textbf{$g$}	& \textbf{$r_+^C$}	& \textbf{$T_+^{Max}$}\\
			\midrule
			0.2 & 0.8293  & 0.0599\\
			0.4 & 1.1293  & 0.0477\\
			0.6 & 1.4964  & 0.0373\\
			\bottomrule
		\end{tabularx}
\end{table}

By following the first law of black hole thermodynamics~\cite{Ghosh:2020tgy,Ghosh:2014pga}
\begin{equation}
dM_+=T_+dS_+ + \phi dg,
\end{equation}
we obtain the expression for the entropy of the $4D$ nonsingular EGB black hole as 
\begin{equation} \label{B_s}
S_+=\frac{A}{4} +2\pi \alpha\log\left(\frac{A}{A_0}\right)-2\pi g^3\left(\frac{1}{r_+}+\frac{2\alpha}{3r_+^3}\right),
\end{equation}
where $A=4\pi r_+^2$ is the black hole event horizon area and $A_0$ is a constant with the units of area. It is noteworthy that magnetic charge $g$ and Gauss--Bonnet coupling constant $\alpha$ contribute two extra terms in the expression of entropy leading to the modification in the usual area law~\cite{Bekenstein:1973ur}. Switching off magnetic charge $g=0$, leads to 
\begin{equation}
S_{+}=\frac{A}{4}+2\pi \alpha \log(\frac{A}{A_0}),
\end{equation}
\textls[-20]{entropy of the $4D$ EGB black holes~\cite{Singh:2020nwo}. In case ($\alpha=0$), we obtain the entropy of the Schwarzschild} black holes obeying area law~\cite{Kumar:2020bqf,Ghosh:2020tgy}.

\subsection{Stability Analysis}
 
 Next, we find the local and global stability regions of the $4D$ nonsingular EGB black holes, respectively, via the behavior analysis of the specific heat ($C_+$) and Gibb's free energy ($F_+$) of the black holes.
The local thermodynamic stability of thermodynamic systems is governed by the behavior of their specific heat ($C_+$).  The thermodynamic systems with $C_+>0$ and $C_+<0$, respectively, are locally stable and unstable. We calculate the specific heat of $4D$ nonsingular EGB black holes by using~\cite{Ghosh:2008jca,Ghosh:2014pga}
\begin{equation}
C_+=\frac{\partial M_+}{\partial T_+}\equiv\left(\frac{\partial M_+}{\partial r_+}\right)\left(\frac{\partial r_+}{\partial T_+}\right),
\end{equation}
 which reads
\begin{equation} \label{BAdSCp}
C_+ =-2\pi r_+^2\left[\frac{(1+\frac{g^3}{r_+^3})^2(r_+^2+2\alpha)^2\left(1-\frac{\alpha}{r_+^2}-2\frac{g^3}{r_+^5}(r_+^2+2\alpha)\right)}{r_+^4-\alpha(5r_+^2+2\alpha)-D\frac{g^3}{r_+^3}-E\frac{g^6}{r_+^6}}\right],
\end{equation}
with
\begin{equation*}
D=2\left(5r_+^4+2\alpha(10r_+^2+7\alpha)\right) \quad \text{and} \quad E=2\left(r_+^2+2\alpha\right)^2.
\end{equation*}
By keeping ($g=0$), one obtains the specific heat of $4D$ EGB black holes~\cite{Ghosh:2020tgy}
\begin{equation}
C_+ =-\frac{2\pi r_+^2(r_+^2+2\alpha)^2\left(1-\frac{\alpha}{r_+^2}\right)}{r_+^4-\alpha(5r_+^2+2\alpha)},
\end{equation}
which further in GR limits ($\alpha=0$) becomes, $C_+ = -2 \pi  r_+^2$,
which is the expression of specific heat for the Schwarzschild black hole ~\cite{Cho:2002hq,Kumar:2018vsm}.

\begin{figure}[H] 
		\includegraphics[width=7 cm]{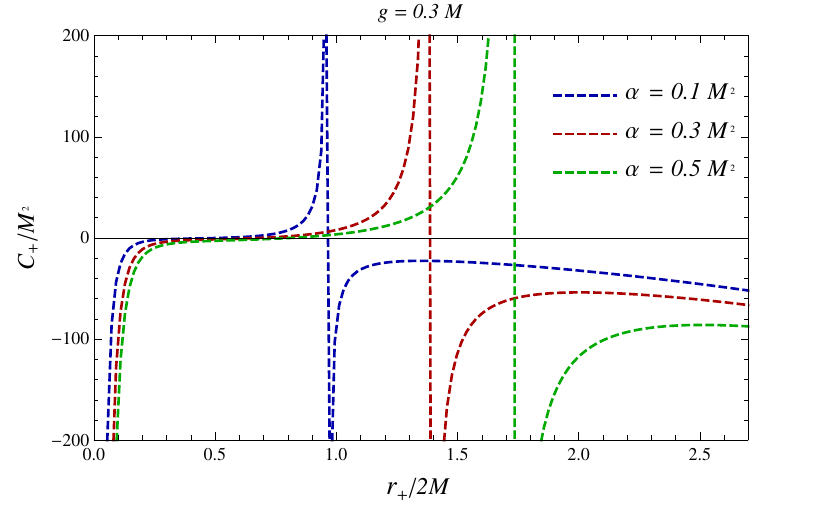}
		\includegraphics[width=7 cm]{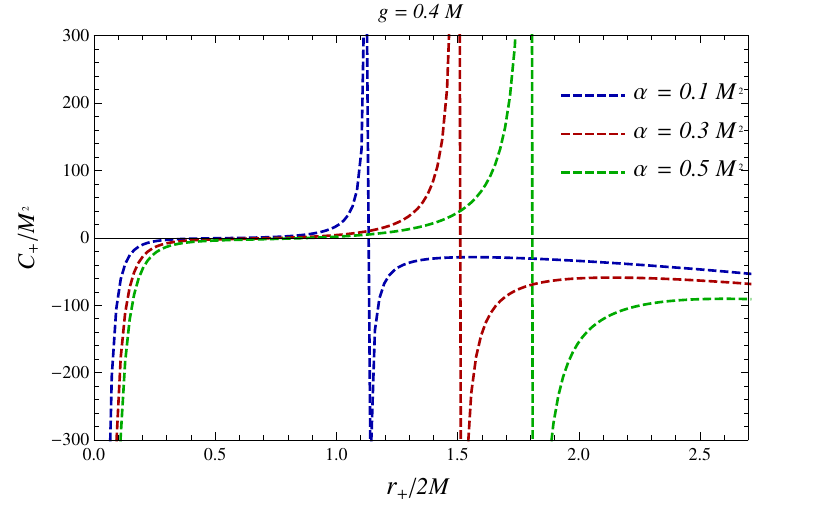}\\
		\includegraphics[width=7 cm]{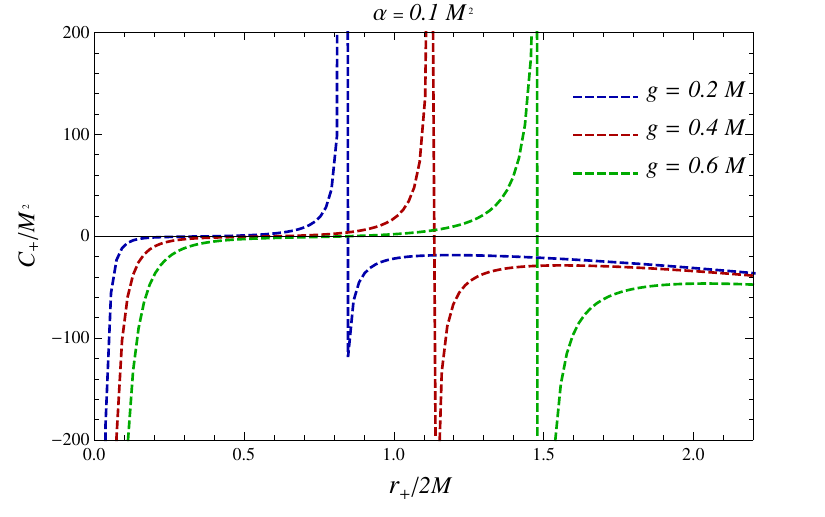}
		\includegraphics[width=7 cm]{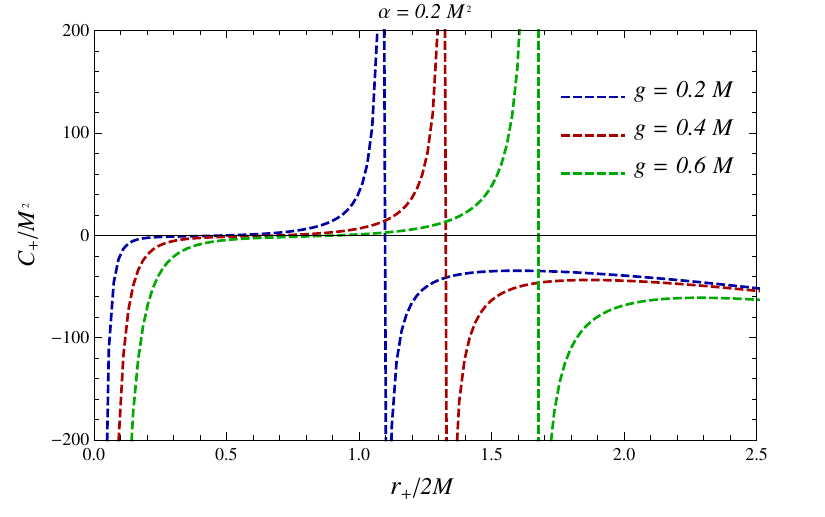}
	\caption{\textls[-15]{$4D$ nonsingular EGB black holes specific heat ${C_+}$ \textit{vs} horizon $r_+$  for different values of $\alpha$ and $g$}.}
	\label{fig:c1}
\end{figure}

We depict the numerical results of the specific heat of $4D$ nonsingular EGB black holes for various values of parameters in Figure~\ref{fig:c1}. It is evident that the black holes with tiny horizon radius, $r_+<r_*$, having negative specific heat ($C_+<0$) are locally unstable. The~diverging specific heat of $4D$ nonsingular EGB black holes flips its sign from positive to negative at critical radius, $r_+^c$, confirming the existence of second-order phase transition~\cite{Davies:1977bgr} between locally stable and unstable black holes. Hence, the black in the region, $r_*<r_+<r_+^c$, are locally stable, whereas the black holes with radius $r_+<r_*$ and $r_+>r_+^c$ are locally unstable (cf. Figure~\ref{fig:c1}). It is also noteworthy that the value of a critical radius $r_+^c$ increases with $g$ and $\alpha$. We note that $r_+^C$ and $r_*$ have larger values when compared with the analogous $5D$ case~\cite{Kumar:2020bqf}.

\vspace{-6pt}

 Further, we are going to examine the global stability of the black hole via Gibb's free energy; the reason for this is that even if the black hole is thermodynamically stable, it~could be globally unstable or vice-versa~\cite{Kumar:2018vsm}. The thermodynamic systems with negative Gibb's free energy ($F_+<0$) are globally stable, whereas, on the other hand, those with positive free energy ($F_+>0$) are globally unstable~\cite{Kumar:2018vsm}. The Gibb's free energy of a black hole can be defined as~\cite{Herscovich:2010vr,Kumar:2018vsm,Cai:2001dz}
\begin{equation}\label{fe}
F_+=M_+-T_+S_+
\end{equation}
to obtain Gibb's free energy for $4D$ nonsingular EGB black holes
\begin{adjustwidth}{-\extralength}{0cm}
\begin{equation}
F_+=\frac{1}{4r_+}\left[2(r_+^2+\alpha)
+\frac{\left(-r_+^2+\alpha+2\frac{g^3}{r_+^3}(r_+^2+2\alpha)\right)\left(3r_+^2+6\alpha\log(\frac{A}{A_0})-2\frac{g^3}{r_+^3}(3r_+^2+2\alpha)\right)}{3\left(1+\frac{g^3}{r_+^3}\right)\left(r_+^2+2\alpha\right)}\right]
\end{equation} 
\end{adjustwidth}
which turns into the expression of free energy of $4D$ EGB black hole~\cite{Singh:2020nwo}, when we take $g=0$
\begin{equation}
F_+=\frac{1}{4r_+}\left[2(r_+^2+\alpha)+\frac{(-r_+^2+\alpha)\left(r_+^2+2\alpha\log(\frac{A}{A_0})\right)}{r_+^2+2\alpha}\right],
\end{equation} 
further, we take GR limits ($\alpha= 0$), to obtain the Gibb's free energy of Schwarzschild black holes~\cite{Kumar:2018vsm,Ghosh:2020tgy,Singh:2020nwo},
\begin{equation}
F_{+}=\frac{r_+}{4}.
\end{equation}

\begin{figure}[H] 
\includegraphics[width=7 cm]{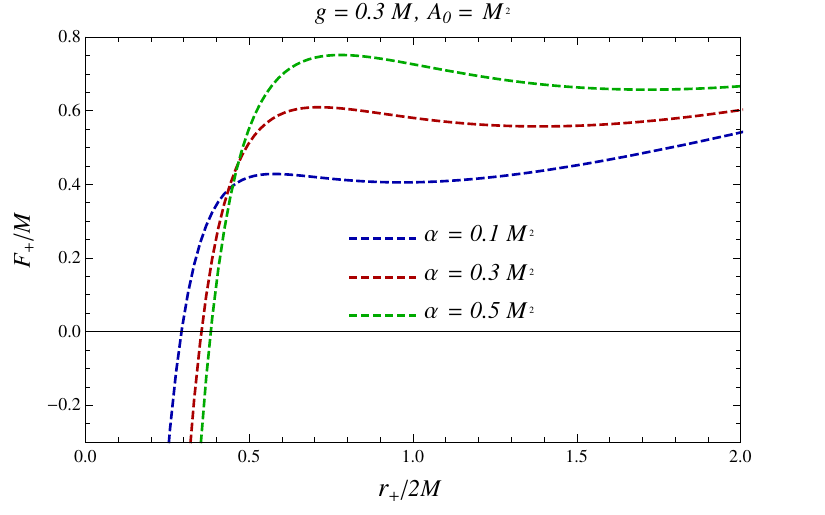}
\includegraphics[width=7 cm]{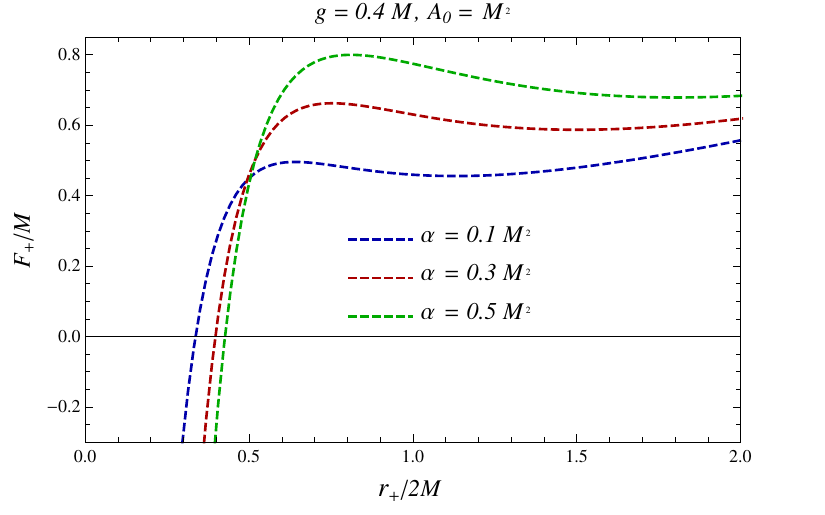}\\
\includegraphics[width=7 cm]{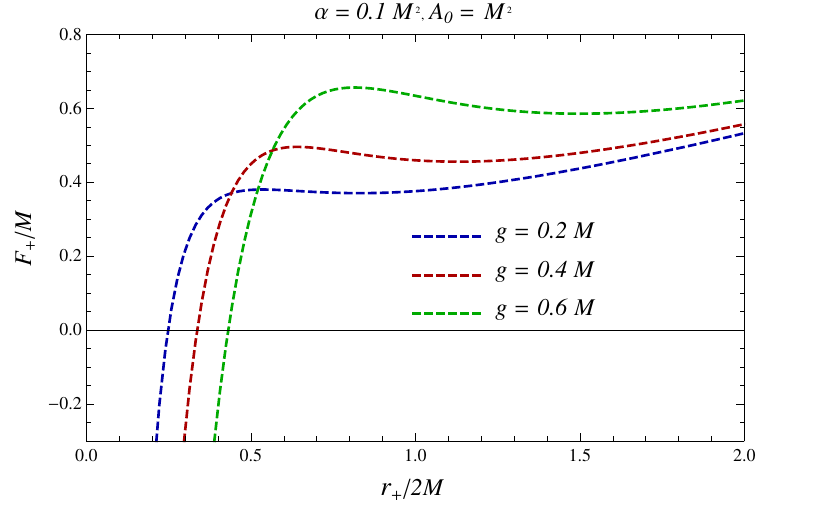}
\includegraphics[width=7 cm]{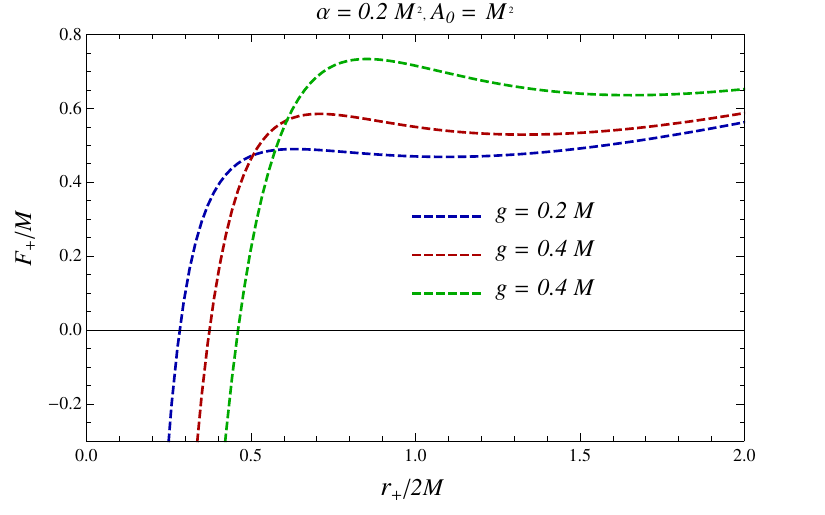}
\caption{$4D$ nonsingular EGB black holes Gibb's free energy $F_+$ \textit{vs} horizon $r_+$ for various values of $\alpha$ and $g$.}
\label{fig:Bf1}
\end{figure}

We plot the behavior of Gibb's free energy with the varying horizon radius of $4D$ nonsingular EGB black holes in Figure~\ref{fig:Bf1}. It is well known that the regions of parametric space with negative Gibb's free energy are preferable because the black holes are globally stable there. Gibb's free energy behavior analysis of $4D$ nonsingular EGB black holes reveals that the black holes with a smaller horizon radius with $F_+<0$ are globally stable, whereas those with a larger horizon radius have positive Gibb's ($F_+>0$) free energy and are globally unstable. It is noteworthy that the black holes with a more considerable value of Gauss--Bonnet coupling constant $\alpha $ or magnetic monopole charge $g$ have a broader region of global stability. Comparing $4D$ nonsingular black holes with the $5D$ case~\cite{Kumar:2020bqf}, we find that the $4D$ nonsingular black holes are globally stable at a larger horizon radius.

\vspace{-3pt}

\subsection{Black Hole Remnant}
\begin{figure}[H]
\includegraphics[width=7 cm]{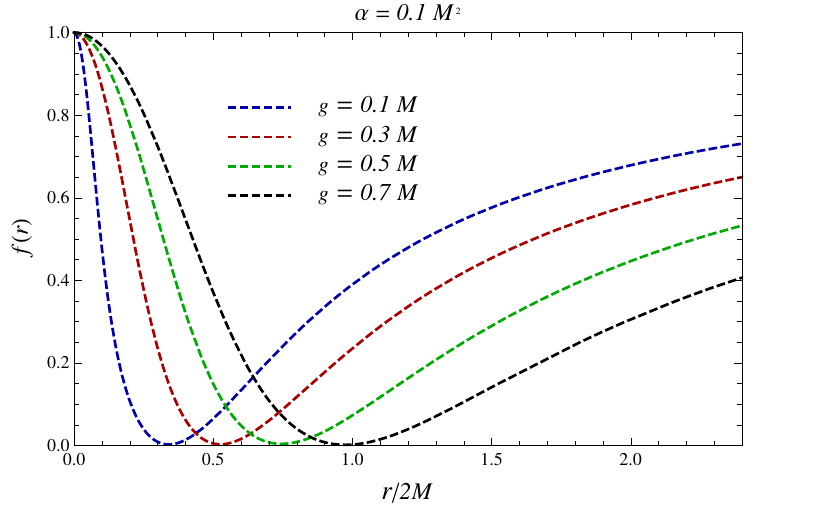}
\includegraphics[width=7 cm]{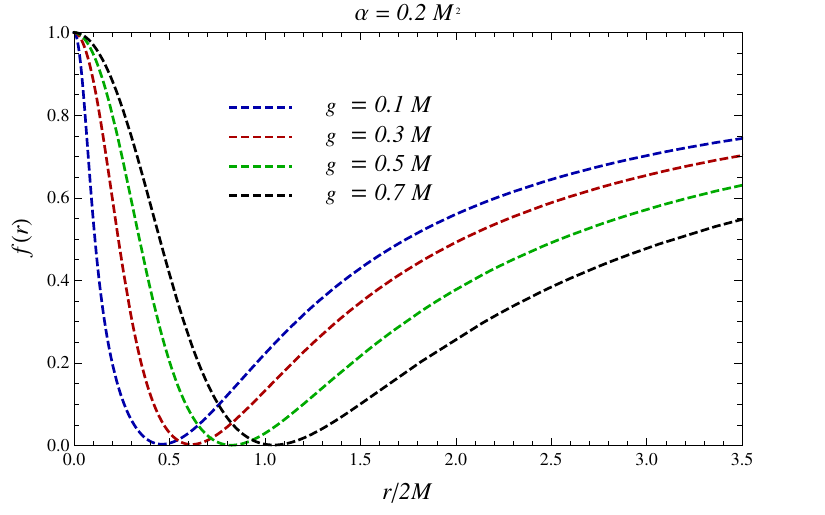}\\
\includegraphics[width=7 cm]{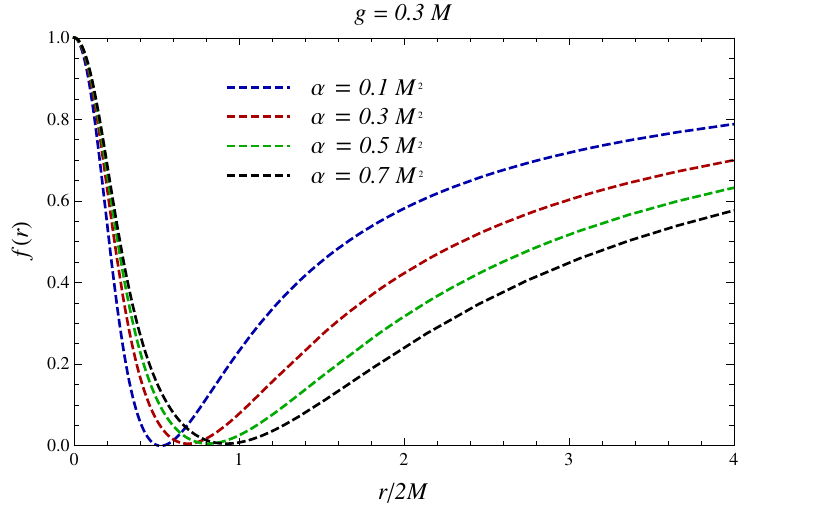}
\includegraphics[width=7 cm]{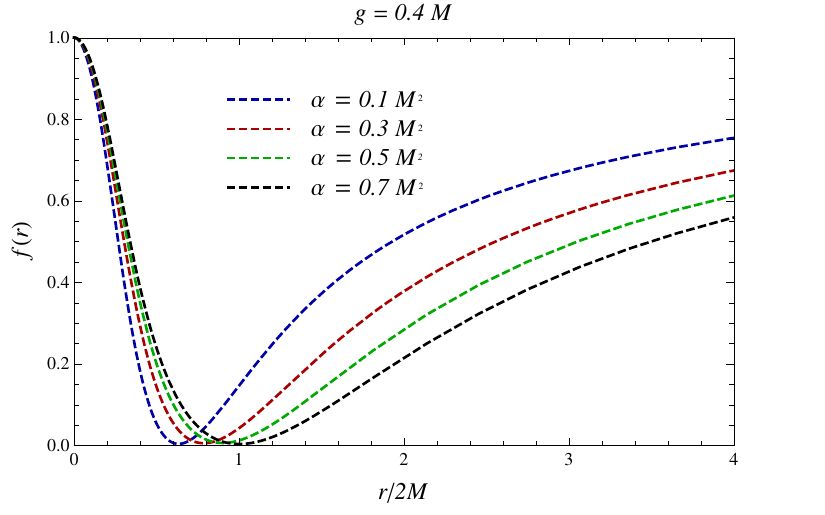}
	\caption{The extremal $4D$ nonsingular EGB black holes Metric function $f(r)$ \textit{vs} $r$  for various values of $\alpha$ and $g$.}
	\label{fig:f3}
\end{figure}

The Hawking evaporation results in either a stable or a long-lived localized stage, known as a black hole remnant. The possibility of the black hole remnant as a source of dark matter~\cite{MacGibbon:1987my} made the study of the black hole remnant fascinating and important. To analyze the emitted feature of $4D$ nonsingular EGB black holes, we plot the metric function (\ref{fr}) vs. varying radius of the extremal $4D$ nonsingular EGB black hole in Figure~\ref{fig:f3}. The numerical analysis of $f^{\prime}(r)|_{r=r_E} = 0$ leads us to find a minimum allowed value, $M_0$ (remnant mass), of black hole mass with corresponding horizon radius $r_0$ (remnant size), for the existence of the black hole solution such that for $M<M_0$, no black hole solution exists. The numerical results of remnant mass ($M_0$) and remnant size ($r_0$) tabulated in Table~\ref{tr2} lead us to conclude that the values of $M_0$ and $r_0$ increase with $\alpha$ and $g$. It turns out that the $4D$ nonsingular EGB black holes have greater remnant mass when compared with nonsingular $5D$ EGB black holes~\cite{Kumar:2020bqf} remnant mass. Hence, the remnant size in the $4D$ case is also larger.

\vspace{-3pt}

\unskip

\begin{table}[H]
\caption{The remnant radius ($r_0$) and the remnant mass ($M_0$) of $4D$ nonsingular EGB black hole for different values of ($\alpha$) and $g$. \label{tr2}}
\newcolumntype{C}{>{\centering\arraybackslash}X} 
\begin{tabularx}{\textwidth}{CCC}
		\toprule
			\textbf{$g = 0.3$}\\
		\midrule
			\textbf{$\alpha$}	& \textbf{$M_0$}	& \textbf{$r_0$} \\
			\midrule
			0.1 & 0.423  & 0.538\\
			0.3 & 0.605  & 0.695\\
			0.5 & 0.745  & 0.817\\
			0.7 & 0.864  & 0.918\\
		\midrule
		\textbf{$\alpha = 0.1$}\\
		\midrule
		\textbf{$g$}	& \textbf{$M_0$}	& \textbf{$r_0$}\\
			\midrule
			0.1 & 0.324  & 0.342\\
			0.3 & 0.423  & 0.538\\
			0.5 & 0.571  & 0.757\\
			0.7 & 0.737  & 0.982\\
			\bottomrule
		\end{tabularx}
\end{table}
    
\section{Discussion}
\label{sec5}
It is believed that general relativity requires modification in the regions where the spacetime curvature becomes high and ultraviolet theory (UV) complete. It has been established that the addition of the higher-order in curvature terms can improve the UV properties of Einstein's gravity. EGB gravity, with quadratic curvature, has several additional better properties than Einstein's general relativity and is also free from ghosts~\cite{Boulware:1985wk}.
Glavan and Lin ~\cite{Glavan:2019inb} reformulated EGB gravity by rescaling of the Gauss--Bonnet coupling constant to $\alpha/(D-4)$ and taking limit $D\to 0$ at the level field equations to make a non-trivial contribution to the gravitational dynamics even in  $D=4$~\cite{Glavan:2019inb}. The~theory preserves the number of degrees of freedom and remains free from Ostrogradsky instability~\cite{Glavan:2019inb}.   Hence, we have obtained an exact  $4D$ nonsingular EGB black hole metric, characterized by three parameters, mass ($M$), the Gauss--Bonnet coupling constant ($\alpha$) and magnetic monopole charge ($g$),  and it regains the $4D$  EGB ~\cite{Glavan:2019inb} metric as a particular case in the absence of a magnetic charge $(g=0).$

The analysis of horizon structure leads us to obtain a maximum allowed value of Gauss--Bonnet coupling constant, $\alpha_0$, for fixed values of $M$ and $g$ such that for $\alpha>\alpha_0$ no black hole solution exists. Whereas $\alpha<\alpha_0$ and $\alpha=\alpha_0$ correspond to black holes with double and single degenerate horizon radii. We computed the Hawking temperature ($T_+$), entropy ($S_+$), heat capacity ($C_+$) and Gibb's free energy ($ F_+ $) associated with the horizon radius of the black hole. In turn, we analyzed in detail the specific heat and found that there existed a second-order phase transition with diverging $C_+$ at a critical radius, $r_+^c$. The black holes with $r_+<r_+^c$ were found to be locally thermodynamically stable, and on the other hand, black holes with $r_+<r_*$ and $r_+>r_+^c$ were found to be locally unstable. While the analysis of Gibb's free energy $F_+$ leads us to find that $4D$ nonsingular EGB black holes with smaller horizon radius' are globally stable with $F_+<0$, whereas those with the more considerable value of horizon radius having $F_+>0$ are globally unstable. Finally, we have shown that the $4D$ nonsingular EGB black holes evaporation results in a stable black hole remnant with zero temperature $T_+=0$ and positive specific heat $C_+>0$. 

Many exciting avenues are available for future work, e.g., it will be intriguing to analyze such black holes in the AdS background, consider rotating counterparts and constraint parameters $\alpha$ and $g$ within the resolution of today's observational facilities. The magnetic charge profoundly influences the horizon radius, which may have several astrophysical consequences, such as wormholes, accretion onto black holes, and gravitational lensing, which are also interesting for future projects. 
\vspace{6pt} 

\authorcontributions{Conceptualization,  S.G.G.; methodology, A.K., D.B. and S.G.G.; software, A.K.; validation, A.K., D.B. and S.G.G.; formal analysis, S.G.G. and A.K.; investigation, S.G.G. and A.K.; writing---original draft preparation, S.G.G. and A.K.; writing---review and editing, S.G.G., D.B. and A.K.; visualization, A.K. and S.G.G.; supervision, S.G.G. and D.B.; project administration, S.G.G. All authors have read and agreed to the published version of the manuscript.}

\funding{This research received no external funding.}
\institutionalreview{Not applicable.}


\informedconsent{Not applicable.}


\dataavailability{Not applicable.} 

\acknowledgments{The authors would like to thank Rahul Kumar Walia for fruitful discussion.}

\conflictsofinterest{The authors declare no conflict of interest.} 

\begin{adjustwidth}{-\extralength}{0cm}
\reftitle{References}

%


\end{adjustwidth}
\end{document}